\documentclass{article}
\usepackage{amsmath}
\usepackage{amssymb}
\usepackage{spconf,graphicx,hyperref}
\usepackage{siunitx, booktabs}

\usepackage[subtle]{savetrees}
\usepackage[font=small, skip=1pt]{caption}

\usepackage{xcolor}
\definecolor{heroblue}{rgb}{0.128,0.37,0.6}
\definecolor{herogreen}{rgb}{0.11,0.496,0.168}
\definecolor{myred}{rgb}{0.7,0.2,0.2}
\newcommand{\heroblue}[1]{\textcolor{heroblue}{#1}}
\newcommand{\herogreen}[1]{\textcolor{herogreen}{#1}}

\title{Generative Audio Extension and Morphing}
\name{Prem Seetharaman$^{\star}$, Oriol Nieto$^{\star}$, Justin Salamon}
\address{Adobe Research, San Francisco, CA, USA}
\begin{document}
\ninept
\maketitle

\renewcommand{\thefootnote}{$\star$}
\footnotetext{Equal contribution.}
\renewcommand{\thefootnote}{\arabic{footnote}} % Resets to numbers for future footnotes

\begin{abstract}
In audio-related creative tasks, sound designers often seek to extend and morph different sounds from their libraries. 
Generative audio models, capable of creating audio using examples as references, offer promising solutions. 
(i) given an audio reference, we can extend it both forward and backward for a specified duration, and (ii) given two audio references, we can morph them seamlessly for the desired duration. 
Furthermore, we show that by fine-tuning the model on different types of stationary audio data we mitigate potential hallucinations. 
The effectiveness of our method is supported by objective metrics, with the generated audio achieving Fréchet Audio Distances (FADs) comparable to those of real samples from the training data.
Additionally, we validate our results through a subjective listener test, where subjects gave positive ratings to the proposed model generations.
This technique paves the way for more controllable and expressive generative sound frameworks, enabling sound designers to focus less on tedious, repetitive tasks and more on their actual creative process.
\end{abstract}
\section{Introduction}
\label{sec:intro}

Sound design is a crucial aspect of various creative fields, including film, television, video games, and virtual reality. It involves the art and practice of creating auditory elements that enhance the narrative, evoke emotions, and provide an immersive experience. 
Sound designers meticulously craft and manipulate sounds to achieve a desired auditory effect, often morphing natural and synthetic sounds to create unique auditory landscapes. This process requires a deep understanding of acoustics, psychoacoustics, and the technical skills to use various audio tools and software~\cite{FoleyGrail21}.
The ability to generate seamless audio \emph{extensions} and \emph{morphs} is of paramount importance for sound designers. 
These practitioners often face the challenge of extending or morphing audio clips to fit specific scenes or transitions without introducing artifacts or unnatural elements, as it is often the case that original footage may have been cut short or started slightly later than desired.
Traditional methods can be time-consuming and may not always yield satisfactory results, especially when dealing with complex soundscapes or non-stationary sounds.

Recent advancements in audio-based diffusion models have demonstrated significant progress in both text-conditional~\cite{StabilityFast,AudioLDM23,DiffSound23} and unconditional~\cite{kong2021diffwave,AudioLDM2} audio generation. 
Furthermore, enhanced control mechanisms for such generative models have been proposed, leveraging audio characteristics for text-based conditions~\cite{AudioGenTextCond2024, sila25} and refining latent representations to achieve fine-grained control via time-varying signals~\cite{Sketch2Sound25, MambaFoley25, tfoley24, syncfusion24}.
Audio extension has also been explored for general audio~\cite{greshler2021catch}, and particularly within the domain of music~\cite{ControllableMusic23}, while audio morphing techniques, often referred to as audio \emph{in-painting}, have been applied to generic audio scenarios~\cite{AudioX25, MorphFader25, specMaskGit2024, soundmorpher24, DeepAudioInpainting24, simGraphsInpainting18}, as well as music~\cite{gacela21, MorphingMusicalSounds21}.
Nevertheless, these methodologies do not specifically address the requirements of sound designers, frequently resulting in suboptimal outcomes or models that are incompatible with the types of sounds they typically work with, such as special effects and environmental sounds.

Our work addresses these challenges by introducing models capable of producing high-quality, 48kHz, stereo audio extensions and morphs from one audio to another.
By leveraging Diffusion Transformers~\cite{DiT2023} operating on audio latents, we propose a novel latent masking technique combined with a variant of Classifier-Free Guidance~\cite{CFG2021}, resulting in seamless and effective audio extensions and morphs. 
Additionally, we mitigate potential hallucinations in stationary generations by implementing an innovative fine-tuning strategy. 
We evaluate our results using the Fr\'echet Audio Distance~\cite{FAD19}, demonstrating that the generated audio is often indistinguishable from their respective audio prompts in terms of audio quality.
Furthermore, we confirm our findings through a subjective listener test, where participants rate the model's generated audio positively.
With this novel model, our aim is to enhance the creative workflow of sound designers and reduce the manual, tedious efforts required in their tasks.

% techniques in machine learning and audio processing, we aim to provide sound designers with powerful tools that enhance their creative workflow and reduce the manual effort required in their tasks.

The main contributions of our work are as follows:
\begin{itemize}
    \item A model designed to generate extensions and morph between one or two audio prompts, facilitating seamless continuation and merging of audio clips.
    \item A novel Audio Prompt Guidance (APG) technique inspired by Classifier-Free Guidance, which improves the quality and coherence of generated audio.
    \item A novel approach to addressing hallucinations by fine-tuning the model using a synthetic Noise Floor Dataset, ensuring the generated audio remains faithful to the input prompts.
\end{itemize}

% \section{Audio Latent Diffusion Models}
% \label{sec:background}

% - Background on latent diffusion models

% - References:

% - AudioLDM~\cite{AudioLDM23}

% - Stable Diffusion~\cite{StabilityFast}

% - Immerse Diffusion~\cite{immerse-diffusion25}

\section{Method}
\label{sec:method}

\begin{figure*}[t]
  \centering
  \centerline{\includegraphics[width=0.85\linewidth]{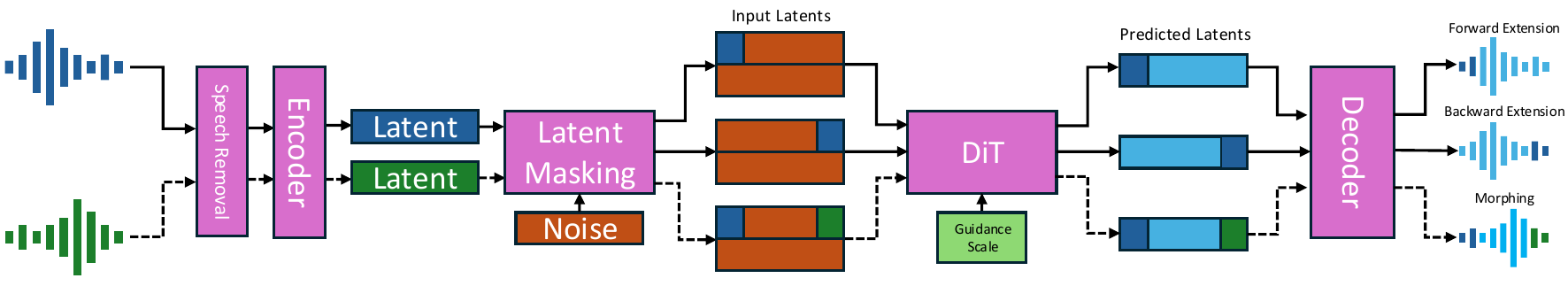}}
  \caption{Proposed block diagram of Generative Extend (solid lines) and Morphing (dashed lines). Two audio examples are shown: the \heroblue{blue} one is used for Generative Extend (both forward and backward) and both the \heroblue{blue} and the \herogreen{green} ones are used for Morphing (from \heroblue{blue} to \herogreen{green}).}
  \label{fig:herofig}
  \vspace{-10pt}
\end{figure*}

The proposed method for Generative Extend and Morphing is depicted in Figure~\ref{fig:herofig}. 
In this section, we describe the entire process, starting from the shared pipeline (inlcuding pre- and post-processing and audio prompt guidance), and then describing each of the two applications individually.

\subsection{Input Preprocessing}

Our primary audience comprises sound designers who specialize in stationary and non-stationary special effects and environmental sounds, so no speech data is employed during training.
Thus, speech components are removed from the audio using~\cite{Soundlift24}.\footnote{This is a well established technique and its authors shared their code.}
Subsequently, the speech-free \emph{audio prompt} $\mathbf{x} \in \mathbb{R}^{T}$ is encoded into a latent representation $\mathbf{z} \in \mathbb{R}^{n \times T_z}$, where $T$ is the number of audio samples, $n$ is the number of channels of the latent, and $T_z$ is the number of frames required to encode $T$ audio samples in the latent space (i.e., $T_z \ll T$).
Any autoencoder architecture capable of generating highly compressed, high-quality latent representations is suitable for this purpose, such as Variational Autoencoders (VAEs)~\cite{VAE2014}, Residual Vector Quantization (RVQ)~\cite{RVQ2027}, or any pre-trained model like the Descript Audio Codec (DAC)~\cite{DAC2023}.

Assuming that we want to generate $d_z$ frames of audio, we then feed the Gaussian noise $\mathbf{z}_G \sim \mathcal{N}(\mathbf{0}, \mathbf{I}) \in \mathbb{R}^{n \times d_z}$ to a Diffusion Transformer (DiT)~\cite{DiT2023} $f^\theta$.
We opt for a DiT architecture due to the availability of an in-house implementation, though the approach is be compatible with other diffusion architectures (e.g., U-Net~\cite{UNet2015}).
Note that $d_z$ is independent of $T_z$, as the amount of generated content $d_z$ does not depend on the length of the audio prompt $T_z$.
Such noise is \emph{masked} using the input prompt latents via $f_M(\mathbf{z}_G, \mathbf{z})$, where $f_M: \mathbb{R} \times \mathbb{R}  \to \mathbb{R}$ is the application-dependent masking function described in following subsections.
The input to the DiT is \emph{both} the masked ($f_M(\mathbf{z}_G, \mathbf{z})$) and unmasked ($\mathbf{z}_G$) noisy latents.
This is depicted in the ``Input Latents" examples in the center of Figure~\ref{fig:herofig}, where the total length of the input latents (orange part including the prompt) is $d_z$ and the length of the prompts (the blue or green small rectangles) is $T_z$.
The DiT \emph{does not denoise} the masked parts of the  noisy latents, thus forcing the generation to align with the original latents.
Note that this setup requires $T_z < d_z$, i.e., the audio prompt needs to be shorter than the total noise fed into the DiT.

% \begin{equation}
% \begin{bmatrix}
% \mathbf{\hat{z}_0} \\
% \mathbf{\hat{z}_1}
% \end{bmatrix} = \begin{bmatrix}
% f_M(\mathbf{z}_G, \mathbf{z}) \\
% \mathbf{z}_G
% \end{bmatrix}    
% \end{equation}

% where $f_M: \mathbb{R} \times \mathbb{R}  \to \mathbb{R}$ is the application-dependent masking function described below, and $\mathbf{\hat{z}}_i \in \mathbb{R} ^ {2n \times d_z}$ is the input to the DiT.

\subsection{Audio Prompt Guidance}
\label{subsec:apg}

To ensure that the generated audio aligns more faithfully with the audio prompt, we modify the DiT so that it applies a variant of \emph{classifier free-guidance} (CFG)~\cite{CFG2021} that we call \emph{audio prompt guidance} (APG).
APG guides the generation towards the distribution of the masked latents and away from the one of the unmasked latents.
This makes the model produce much higher quality audio, with better adherence to the prompt.
Formally:

\begin{equation}
\mathbf{z'} = f^\theta(\mathbf{z}_G) + \gamma\left(f^\theta(f_M(\mathbf{z}_G, \mathbf{z})) - f^\theta(\mathbf{z}_G)\right)
\label{eq:cfglatents}
\end{equation}

where $\mathbf{z'} \in \mathbb{R}^{n \times d_z}$ is the APG output of the DiT and $\gamma$ is the guidance scale.

This approach mirrors the original CFG formulation but is uniquely applied to the same modality as the target data, i.e., audio.
To the best of our knowledge, this is the first work to apply CFG in this manner.

\subsection{Output Postprocessing}

Output $\mathbf{z'}$ is then post-processed by applying the same $f_M$ function used in the input.
Thus, the final post-processed output contains the original prompt latent information (i.e., the audio to be extended/morphed): $\mathbf{\hat{z}'} = f_M(\mathbf{z'}, \mathbf{z})$.

We finally pass this through the respective decoder to obtain the final output audio $\mathbf{y} \in \mathbb{R}^{d}$, where $d$ is the total duration in audio samples.
Note that the total number of generated samples in the post-processed output will be $d - T$ for Generative Extend and $d - 2T$ for Morphing, as the raw APG output of the DiT will be replaced by the original prompt information.
This is shown in Figure~\ref{fig:herofig}, in the final generated audio waveforms on the right, where the total length of the final full audios is $d$, the length of the prompts (dark blue or green) is $T$, and the length of the postprocessed generated audios (light blue) is $d - T$ and $d - 2T$ for extend and morphing, respectively.

\subsection{Generative Extend}
\label{subsec:genextend}

The application of Generative Extend can be applied in two different modes: \emph{forward} and \emph{backward}.
In both cases, a single audio prompt $\textbf{x}$ is required.
For the forward extension (top example on Figure~\ref{fig:herofig}), the masking function $f_M$ copies the input latent information into the \emph{beginning} of the noisy/output latent.
In this way, the prompt is extended forward across the time dimension.
On the other hand, the masking function $f_M$ during the backward extension (middle example of Figure~\ref{fig:herofig}) copies the input latent information to the \emph{end}, so that the audio prompt is extended backwards.
The final output will have a duration of $d$ audio samples, $T$ of which will be the input query samples.

% Note that it is possible to mask both the beginning \emph{and} end of the noisy latents, thus achieving what is referred to as \emph{audio in-painting}.
% A similar idea is what we describe below as Generative Morphing.

\subsection{Generative Morphing}
\label{subsec:genblend}

We propose using this framework to generate audio information to \emph{morph} from audio prompt $\textbf{x}_1$ to $\textbf{x}_2$ in a specific time (example depicted at the bottom of Figure~\ref{fig:herofig}).
This can be seen as audio in-painting with potentially very different initial and ending audios.\footnote{
Note that this is different than \emph{timbral} morphing, where two sounds are converted into a new sound without temporal transitions.}
To do so, we need to apply the masking function $f_M$ twice: once to replace the noisy/output latents at the beginning with audio query $\textbf{x}_1$ and a second time to mask the end of the noisy/output latents using the audio query $\textbf{x}_2$.
Assuming that both $\textbf{x}_1$ and $\textbf{x}_2$ contain $T$ audio samples, the generated morph will be $d - 2T$ audio samples long. 
Note that one could mask the two query latents closer together to adjust for the final desired duration.
Finally, audio prompts can be of any length as long as their total sum is less than the fixed generation length $d$, i.e., $2T < d$.

\section{Training Strategy}

We train a stereo audio encoder and alter the standard DiT training process to achieve the desired Generative Extend and Morphing features.
In this Section, we describe the details of such a process.

\subsection{Stereo Encoding}

We train a VAE with a novel extension towards encoding arbitrary stereo audio of all audio domains.
Its architecture is fully convolutional and is adapted from DAC~\cite{DAC2023}. 
It compresses 48kHz mono or stereo audio into a 256D latent space at 40Hz, replacing residual vector quantization with a KL-regularized continuous latent space.
% Off-the-shelf variational autoencoders for audio (e.g., DAC~\cite{DAC2023}) are either built for single-channel audio, or specifically for music.
Our model can encode stereo audio of all types, with great care taken for accurate spatial positioning.
It does so by parametrizing audio waveforms into mono (the sum of left and right channels) and side (the difference of left and right channels) when encoding the audio into the latent space.
To train the autoencoder, we also apply this parametrization when computing reconstruction loss (e.g. difference between waveforms, spectrograms, etc). 
Finally, we apply extensive data augmentation to ensure we cover all possible spatial widths and positioning.

\subsection{Training the DiT from Scratch}
\label{subsec:training}

We freeze the stereo encoder/decoder and train the DiT from scratch, setting the full generation duration to 13 seconds (i.e., $d = 13 \cdot 48k$), which ensures the model fits within GPU memory constraints while maintaining a practical batch size for training efficiency.

We implement four key modifications during the standard DiT training process to achieve the desired generative capabilities: (i) \emph{Randomly set the duration of the masks} to enable the model to learn to extend/morph audios irrespective of their length; (ii) \emph{Randomly mask the beginning, end, or both ends} of the embeddings. This ensures the model comprehends how to extend/morph in a single training process;
(iii) \emph{Dropout the masking process}, which trains the model with and without masking, thus enabling the effective application of the APG strategy during inference;
and (iv) \emph{CFG on text conditions}:  we train the model with pairs of audios and their corresponding free-form text descriptions, and randomly omit these text conditions. Consequently, the model generates both unconditional and conditional audio during training, with masked and unmasked noisy latents.

\subsection{The Noise Floor Dataset}
\label{subsec:noisefloor}

The model trained on our primary datasets exhibits a tendency to hallucinate, particularly when prompted with relatively stationary sounds such as ambient noise, room tone, and low levels of white noise. 
When this occurs, the output often contains abrupt, sharp sounds that are not present in the audio prompt.
We hypothesize that this phenomenon arises due to a bias in our training data towards short, non-stationary sounds (e.g., footsteps, explosions).
To address this issue, we propose fine-tuning the model using stationary noise data to ensure the model respects the stationary nature of the input and reduces the likelihood of hallucinating non-stationary sounds during stationary extensions or morphs. 
The goal is to reduce hallucinations while producing non-stationary sounds when the prompt contains such data.
To achieve this, we synthesize a new dataset focused on \emph{stationary} sounds, termed the Noise Floor Dataset, for fine-tuning the model.

The Noise Floor Dataset is a novel set containing 1.3M hours of noise floor data. 
Such set requires two key components: (i) \textbf{Room Tone}: This originates from the LibriVox dataset~\cite{LibriVox}, which we pass through~\cite{Soundlift24} to remove all speech and keep only the background sound, i.e., room tone. Our room tone data are composed of 408 audio files with a total duration of 115 hours; and (ii) \textbf{White Noise}: This can be generated on-the-fly, with the desired final length $n$ of the audio file to be created.

Once we have access to these two components, we create the Noise Floor Dataset as follows: we sample $d$ contiguous audio samples from a randomly selected file of the Room Tone data. 
We convolve this audio with on-the-fly generated white noise of the same length to obtain white noise that matches the frequency response of the room tone, thereby synthesizing $d$ new and unique samples containing noise floor. 
To obtain stereo room tone, we simply repeat the last step for each of the two channels. We repeat the entire process for the desired number of files.
The Noise Floor Dataset contains a total of 100k files.

\subsection{Fine-Tuning: Reducing Hallucinations}
\label{sub:hallucinations}

The final step to mitigate hallucinations is to fine-tune the model with the Noise Floor Dataset. 
To achieve this, we continue training the model using this newly synthesized data exclusively. 
We fine-tune the model by asking it to generate forward/backward extensions (i.e., no morphing), as this seemed to yield the best performance.

We experiment with different numbers of fine-tuning iterations: 10k, 15k, and 20k.
The qualitative results suggest that the higher the number of iterations, the less the model hallucinates, as expected. 
However, increasing the number of iterations leads to a noticeable reduction in prompt fidelity, potentially suggesting a mild effect of catastrophic forgetting.
Thus, we aim to find a balance between reducing hallucinations and maintaining faithfulness to the original prompt. 
Qualitative evaluation indicates that 10k iterations tend to yield the best results for mono models, while 20k iterations are optimal for stereo models.

\section{Experiments}
\label{sec:experiments}

In this section, we describe the experiments conducted to evaluate our model, both objectively and subjectively. Additionally, we strongly encourage readers to listen to output examples of our models on our companion website.\footnote{\url{https://urinieto.github.io/genextend_html}}

\subsection{Experiments Setup}
\label{subsec:expsetup}

% We make use of an internal dataset composed of roughly 1M audio files to train the models in our experiments.
% Such dataset contains around 7.5k hours of sound effects, environmental sounds, ambient, etc., around 75\% of which is professionally recorded with no background noises.
% We use an internal dataset of 1M audio files ($\approx$7.5k hours) to train our models. Approximately 75\% of the data consists of professionally recorded, noise‑free sound effects, environmental sounds, and ambient recordings, targeted for sound designers.
% We make use of an internal dataset composed of approximately 1M audio files for training. 
% The audio includes sound effects, environmental sounds and ambiences, most of which were professionally recorded. 
We combine proprietary sound‑effect datasets with CC‑licensed general‑audio datasets, totaling 1.1M labeled audio samples. Sound‑effect data usually contain a single clean event, while general‑audio data are noisier and may include multiple, underspecified events.
The dataset does not contain music or speech, which were out of scope for the intended application.
Additionally, we augment the metadata of such set by using Mixtral 8x7B~\cite{mixtral_2024} to produce free-form text descriptions, resulting in a set with reliable audio-text pairs.
The audio of our entire set is sampled at 48kHz, and it is mostly stereo.
We randomly downmix to 1 channel during training so that our model will be able to produce both stereo and mono generations.
We use a set of 98 carefully curated stereo audio clips for evaluation.
Such clips originate from professional sound design testers, aiming to cover all realistic cases of the applicability of this model.
There is no overlap between the 1M set described above and this set.
We employ such evaluation set on all of our experiments, unless stated otherwise.

% - Dataset:

	% - 1,058,238 audio files
    % - 7.5k hours of audio
    % - Mixtral augmentations

Our DiT architecture has 8 heads and 24 layers with SwiGLU activations~\cite{swiglu20}, and cross-attends to text conditions only on the first layer.
We use v-prediction~\cite{vPrediction22} for our diffusion process on 13s ($d = 624k$) audio latents, denoised in 24 steps, and set our audio prompt guidance scale to $\gamma = 5$ unless otherwise specified.
We train it with an AdamW optimizer on a standard MSE loss on the latents with a learning rate of 1e-4 and weight decay of 1e-2.
We use a linear warmup of 4k steps and a cosine decay schedule with a 0.5 scale.
We train for a total of 400k iterations, applying 10\% dropout to the embeddeding and attention layers.
We apply Exponential Moving Average to model weights every 100 steps, with 0.99 decay.
We train it on 32 A100 40GB GPUs on the 1M dataset described above, with an effective batch size of 256 (i.e., 3,328 seconds per batch).
We fine-tune it on 8 A100 40GB GPUs using the Noise Dataset described in Section~\ref{subsec:noisefloor} for 15k iterations.
Finally, the latents are randomly masked either at the beginning, end, or both ends, and we dropout such masking by 50\%.
The length of the masks is sampled form $\mathcal{U}(0, 3.25)$ (in seconds).
The text input is dropped (i.e., train unconditionally) 20\% of the times.

\subsection{Evaluation Metrics}
\label{subsec:evalmetrics}

We utilize the standard Fr\'echet Audio Distance (FAD)~\cite{FAD19} to objectively measure the quality of our results. This metric relies on two key attributes: (i) a reference set to compare the distribution of the target audios, and (ii) an audio model that can embed the audios into a compressed yet semantically meaningful space. 
For the reference set, we use the Audition SFX dataset~\cite{adobe_audition_sfx}, which contains approximately 10k high-quality sound effects. We choose CLAP~\cite{laionclap2023} as the embedding model because it can operate on 48kHz stereo files. When evaluating generated extensions and morphs, we only use the actual generated content, ensuring that no leakage from masked latents affects the FAD computation. Since FAD is an actual audio distance metric, lower scores indicate better quality.
% We employ the \texttt{fad\_pytorch} package to compute the FAD scores.\footnote{\url{https://github.com/drscotthawley/fad_pytorch}}
Finally, for the subjective evaluation at the end of this section, we use the standard Mean Opinion Score (MOS).

\subsection{Objective Audio Quality Results}
\label{subsec:audioquality}

The objective audio quality experiment results are shown in Table~\ref{tab:audioquality}. 
As baselines, we compute FAD for the original audio (i.e., the audio prompts used for the generation), 1k files randomly sampled from the Noise Floor dataset described in Section~\ref{subsec:noisefloor}, and pure White Noise.
Additionally, we implement a classical yet robust approach for extension based on~\cite{MCDERMOTT2011926}, and denoted ``Convolutional Noise Matching.''
This approach convolves white noise with the envelope of the input signal, allowing to reproduce the room tone/noise floor of the prompt.
We could not identify other relevant open-source approaches that tackle this problem specifically (e.g.,~\cite{floresgarcia2023vampnet} only extends/morphs music and~\cite{Lugmayr2022RePaint} images;~\cite{MorphFader25} only does timbral morphing --not temporal; etc.).

As expected, White Noise and Original Audio set the lower and upper performance bounds, respectively.
Convolutional Noise Matching yields similar results as noise floor, which is expected given the nature of such approach, targeted for stationary sounds.
Interestingly, Generative Extension and Morphing produce good scores, close to the original prompts. 
The morph is almost indistinguishable from the audio prompts (0.432 vs 0.426), indicating that the audio quality is comparable to the actual prompts used. 
The Noise Floor, representing synthesized ambient sound, obtains worse FAD results, as expected, due to the absence of non-stationary sounds in this set.

\begin{table}[tb]
\footnotesize
\centering
\caption{Objective audio quality evaluation. 
% We employ three different baselines: White Noise (synthetically generated), Noise Floor (1k files sampled from our novel dataset), and Original Audio (the audio prompts used for the generations).
}
\label{tab:audioquality}
\sisetup{
    reset-text-series = false, 
    text-series-to-math = true, 
    mode=text,
    tight-spacing=true,
    round-mode=places,
    round-precision=2,
    table-format=2.2,
    table-number-alignment=center
}
\begin{tabular}{l c}
    
    Approach & {FAD $\downarrow$}\\
    \toprule
    GenExtend &            0.520 \\
    GenMorph  &            \bfseries 0.432 \\
    Convolutional Noise Matching~\cite{MCDERMOTT2011926}  & 0.599 \\
    % \midrule
    % AudioX~\cite{AudioX25}  & \bfseries -1 \\
    \midrule
    White Noise & 1.358 \\
    Noise Floor &    0.586 \\
    \midrule
    Original Audio  & \bfseries 0.426  \\
\end{tabular}
\end{table}

% \subsection{Duration Ablation}
% \label{subsec:duration}

% Use different extension durations during training and see impact on FAD

\subsection{Audio Prompt Guidance Ablation}
\label{subsec:cfg}

\begin{figure}[tbh]
  \centering
  \centerline{\includegraphics[width=0.9\columnwidth]{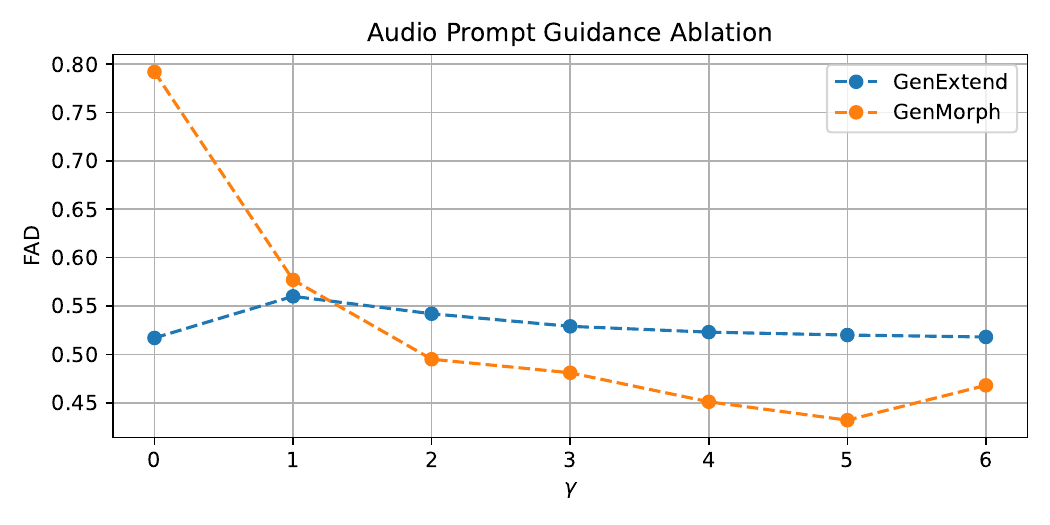}}
  \caption{Ablation of the Audio Prompt Guidance technique. 
  % The higher the guidance scale $\gamma$, the more guided the generation will be towards the masked noisy latents and away from the unmasked ones.
  }
  \label{fig:apgablation}
\end{figure}

On Figure~\ref{fig:apgablation}, we present the results of the ablation study on the proposed Audio Prompt Guidance. We vary the guidance parameter $\gamma$ from 0 (no guidance) to 6 (strong guidance towards the masked noisy latents). As shown, higher $\gamma$ values improve the results in terms of FAD. However, at $\gamma=6$, the scores decline for Generative Morphing, while remaining stable for Generative Extend. Therefore, we select $\gamma=5$ as the optimal guidance, achieving high-quality generations that closely follow the audio prompts.

Interestingly, no guidance towards masking latents seems to be particularly beneficial for morphing, but not for the extension.
We hypothesize that, while the generation does not faithfully follow the masked prompt, the quality of it is quite high, as it has to stay in the same distribution.
On the other hand, without guidance, generative morphing is not capable of producing two potentially different sounds to be morphed, therefore making APG crucial during inference.

% \subsection{Noise Finetuning Ablation}
% \label{subsec:noisefloorablation}

% - Report FAD when using Silence fine-tuning vs without

% \begin{table}[t]
% \centering
% \caption{Noise Floor Ablation}
% \label{tab:noisefloorablation}
% \sisetup{
%     reset-text-series = false, 
%     text-series-to-math = true, 
%     mode=text,
%     tight-spacing=true,
%     round-mode=places,
%     round-precision=2,
%     table-format=2.2,
%     table-number-alignment=center
% }
% \begin{tabular}{l c c}
    
%     Approach & \#iters & {FAD $\downarrow$}\\
%     \toprule
%     GenExtend &  0    &      0.444 \\
%               &  10    &      -1 \\
%               &  15    &      0.520 \\
%               &  20    &      -1 \\
%     \midrule
%     GenBlend  &    0   &        -1 \\
%               &  10    &      -1 \\
%               &  15    &      0.423 \\
%               &  20    &      -1 \\
    
% \end{tabular}
% \end{table}

\subsection{Listener Study}
\label{sub:listenerstudy}

We conducted a listener study to subjectively assess the quality of one of our proposed tasks: Generative Extend. 
The study involved 15 participants, including 12 professionals well-versed in video editing workflows and 3 audio/video researchers. 
Participants were asked to rate extensions on a scale from 1 (worst) to 5 (best) across three dimensions: Smoothness --how smooth the transition is from audio prompt to extension; Consistency --how semantically consistent the extension is with the prompt; and Quality --how good the extension sounds overall.
On average, each participant rated 100 audio extensions, sampled from a dataset of 298. 
In total, 1,435 ratings were collected.

The results, presented in Table~\ref{tab:listenerstudy}, include both forward and backward extensions. 
The highest average score was attributed to Consistency, with a Mean Opinion Score (MOS) of 3.8. 
However, the MOS for Smoothness and Quality were only slightly lower, at 3.5, indicating positive perceptions of these characteristics as well. The median score of 4 was assigned to all three audio characteristics, corresponding to the descriptors: ``Quite Seamless," ``Quite Related," and ``Good" for Smoothness, Consistency, and Quality, respectively. 
These encouraging results highlight the significant potential of our model for real-world applications.

\vspace{-0.25cm}
\begin{table}[t]
\footnotesize
\centering
\caption{Listener Study: Percentage of subjects per score (1-5) for each audio dimension and average score.
% for each one of the evaluated characteristics of the generated extension, including the percentage of responses for each score (1 is worst, 5 is best).
}
\label{tab:listenerstudy}
\sisetup{
    reset-text-series = false, 
    text-series-to-math = true, 
    mode=text,
    tight-spacing=true,
    round-mode=places,
    round-precision=2,
    table-format=2.2,
    table-number-alignment=center
}
\begin{tabular}{c c S[round-precision=1,table-format=2.1]S[round-precision=1,table-format=2.1]}
    
    Scores $\uparrow$ & {Smoothness} & {Consistency}& {Quality}\\
    \toprule
    1      &      5.9\%      &     2.1\%       &    3.8\%     \\
    2      &      9.1\%      &     2.8\%       &     6.6\%    \\
    3      &      27.5\%     &      13.6\%     &      26.5\%   \\
    4      & \bfseries 36.9\%  &  \bfseries 54\% &   \bfseries  41.1\%    \\
    5      &      20.6\%     &       27.5\%       &     22\%    \\
    \midrule
    Average &  3.5    &      3.8 &           3.5 \\
    
\end{tabular}
\end{table}
% \vspace{-1cm}

% \subsection{Looping Ablation?}
% \label{subsec:looping}

% - What if we use generative blend with the same audio reference?

% - Report FAD

% \subsection{Text Conditioning Results?}
% \label{subsec:textcond}

% - Experiment with prompting text too?

\section{Conclusions and Future Work}
\label{sec:conclusions}

We presented a novel approach for generating high-quality audio extensions and morphs using Diffusion Transformers and a variant of classifier-free guidance that we call \emph{audio prompt guidance}. 
Our method effectively addresses the challenges faced by sound designers, providing seamless and coherent audio generations that closely follow the input prompts. 
The objective and subjective evaluations demonstrate the effectiveness of our approach, with generated audio achieving Fr\'echet Audio Distances comparable to original audio and receiving positive ratings from sound design experts.
For future work, we aim to explore several directions to further enhance our model:
(i) investigate the application of our method to other types of audio data, such as speech and music;
(ii) integrate additional control mechanisms to allow sound designers to specify more detailed characteristics of the generated audio, such as specific timbres or dynamic ranges;
(iii) investigate fine-tuning pre-trained generative models with our method, rather than training them from scratch; and
(iv) apply distillation or other recent techniques such as Flow Matching~\cite{lipman2023flow} or Inductive Moment Matching~\cite{imm_2503_07565} to help reduce the number of diffusion steps during inference.

By pursuing these directions, we hope to create even more powerful and versatile tools for sound designers, enabling them to focus on their creative processes and produce high-quality audio content with greater ease and efficiency.

\footnotesize
\bibliographystyle{IEEEbib}
\bibliography{refs}

\end{document}